# New dosimetry planning strategy based on continuous dose gradient used to reduce dose estimation errors due to respiratory motion in breast radiation therapy

Mazen Moussallem[1, 2, 3,*], Pauline Harb[4], Hanane Rima[1, 2], Zeina Al Kattar[4], Saad Ayoubi[1, 2]

[1]Radiation Oncology Department, Centre de Traitement Médical du nord, Zgharta, Lebanon;
[2]Centre Hospitalier du Nord, Zgharta, Lebanon;
[3]Doctoral School of Sciences and Technology, and Faculty of Public Health, Lebanese University, Tripoli, Lebanon;
[4]Faculty of Sciences, Lebanese University, Hadath, Lebanon

*Corresponding author, mazenphm@hotmail.com

October 19, 2021

**ABSTRACT**

A new strategy for radiation therapy dosimetry planning (RTDP) used to reduce dose estimation errors due to respiratory motion in breast treatment was illustrated and evaluated in this study. On CT data set acquired for breast treatment, six different RTDP tangential techniques were performed: (i) three-dimensional conformal radiotherapy (3DCRT) with physical wedge, (ii) 3DCRT with virtual wedge, (iii) motion management technique (MMT) with physical wedge, (iv) MMT with virtual wedge, (v) 3DCRT with field-in-field, and (vi) intensity-modulated radiation therapy (IMRT) with direct aperture optimization. These anti-motion techniques were considered to generate continuous dose degradation to avoid drastic changes in the delivered doses that involve the edge regions of the beams. A comparison was made between the delivered and simulated doses, with and without the presence of motions simulations. This study demonstrates that techniques without motion management can be affected by motions that can lead to a difference equal to 19 % between the delivered and the planned doses in a point located near to the beams collimators edges, and to a difference up to 3 % on multi-leaf collimators (MLCs) edges (gamma pass rates were 87.9 % for 3%/3mm). These differences were reduced to 11 % and to less than 3 % (gamma pass rates were 100 % for 3%/3mm) when MMT were used. As a conclusion, in tangential techniques motion can affect dose estimations on the MLCs and beams collimators edges. The proposed MMT reduce this effect to estimate the delivered dose accurately.

# INTRODUCTION

A major challenge in radiation therapy (RT) consists of delivering the optimal dose in the tumor, while limiting the irradiation of surrounding healthy tissues (Greenberger et al., 1996). Tumor under-dosing can lead to recurrence, while overdosing of surrounding healthy tissue can cause serious side effects (Perez et al., 2004).

To prevent any unexpected under-dosing in the target or overdosing in the organs at risk, researchers have worked in several fields to increase the precision of radiation therapy dosimetry planning (RTDP). For example, to optimize the determination of tumor contours, a number of techniques have been used to generate RTDP objective target volumes using an automatic algorithm that is able to combine the information of different image modalities (Moussallem et al., 2012; Konert et al., 2015). Furthermore, quality assurance procedures were routinely performed (e.g., verifications of patient position, treatment planning parameters, dose distribution, and dose calculation) after approval of the RTDP by the radiation oncologist and before the first treatment fraction. Particularly, for sophisticated treatment planning techniques (e.g., IMRT: intensity-modulated radiation therapy), pretreatment quality assurance (QA) measurement methods are currently used to verify that the delivered dose distributions on a homogeneous phantom fit the estimated doses (Chang J et al., 2000; LoSasso et al., 2001; Cho, 2018) computed by the treatment planning system (TPS). However, these QA methods are widely used in static mode (without motion simulations) to verify TPS dose calculation without considering the physiological patient motion during the treatment. Therefore, in the current study, pretreatment QA was completed by comparing the expected (planned) to the delivered (linac) dose profiles in the presence of respiratory motion simulation.

For breast cancer RT, breast intrafraction respiratory motion amplitude varies from 0 to 3 cm with significant variability between a patient to another one (Kubo and Hill, 1996; George et al., 2003; Kaneko and Horie, 2012; Zhang et al., 2006; Kinoshita et al., 2008). Consequently, for breasts with high respiratory motion amplitude, the delivered dose differs from the planned one (George et al., 2003; Kaneko and Horie, 2012; Tanaka et al., 2014; Menon and Smith, 2008). This difference can result to an unwanted and unexpected high dose for the lung and the heart. An overdose in the heart can lead to cardiovascular disease (Clarke et al., 2005). To reduce the effect of breathing motion and accurately estimate the delivered dose, several strategies, including motion-encompassing methods, respiratory-gated techniques, breath-hold techniques, forced shallow-breathing methods, and respiration-synchronized techniques have been used (Zagar et al., 2017; Keall et al., 2006; Rochet et al., 2013). However, all of these strategies are relatively complicated; they need many QA procedures and additional respiratory motion management devices and procedures to be implemented (Berbeco et al., 2005) that a large number of patients cannot tolerate (Keall et al., 2006). Moreover, patients receive an additional unwanted dose based on real-time images used for respiratory management by some devices (Liu et al., 2010; Hugo et al., 2009). All these reasons and, even the high cost of respiratory management devices, limit the use of respiratory motion strategy in RT departments leading to bad dose estimation.

The aim of this study is to introduce and evaluate a new strategy to reduce dose estimation errors induced by respiratory motion in the tangential breast technique based only on dosimetry planning without the use of any additional respiratory motion management device. Thus, RTDP with this MMT can be easily accomplished in any RT department and can be accessible and tolerated by all patients who have a large respiratory amplitude. However, respiratory amplitude of patients must be assessed before preparing the dosimetry planning and this MMT should not be included in the RTDP for patients who have normal respiratory amplitude.

## MATERIALS AND METHODS

*Respiratory breast motion simulation*

According to Kubo and Hill (1996) and George et al. (2003), we assumed that the intrafraction motion of breast patients is only in the anterior-posterior and in the medial-lateral directions (insignificant motion in the superior-inferior patient direction), and this motion follows a pattern similar to that of a sinusoidal curve. Therefore, in a 2D (two-dimensional) plane perpendicular to the beam axis related to a tangential beam, breasts move approximately in a perpendicular direction to the superior-inferior axis of the patient as illustrated by the white arrows in Figures 1 and 2 (as schematized in George et al. (2003) and in Zhang et al. (2006)). Furthermore, Kaneko and Horie (2012) observed that for deep breathing of 50 females in the supine position, the breathing amplitude is equal to $26.13 \pm 8.01$ mm in the sternal angle point, $29.7 \pm 8.25$ mm in the third right rib point, $28.4 \pm 8.38$ mm in third left rib point, $25.44 \pm 7.15$ mm in the xiphoid point, $27.21 \pm 7.81$ mm in the eight right rib point, and $26.57 \pm 7.42$ mm in the eight left rib point. In another study, George et al. (2005) estimated an average period between 2 and 6 seconds for a respiratory cycle, with significant variability of this period during treatment. Consequently, in the current study, all measurements were performed assuming that we have a breast patient with a respiratory cycle amplitude of 2.6 cm and a period of 4 seconds.

*Radiation therapy dosimetry planning techniques*

On left breast clinical computed tomography (CT) scan images, six different RTDP techniques were performed. A commercial TPS, Prowess Panther$^{TM}$ version 5.2 (Prowess, Inc., CA, USA), was used. Prowess Panther TPS has two algorithms; one is fast photon effective (FPE) model that calculates dose based on measured data where all the tissue is assumed to be water with no tissue heterogeneity or effective path length through tissue is taken into account. The other model is collapsed cone convolution superposition (CCCS) that calculates the dose based on full heterogeneity correction and is a full 3D dose calculation. In Prowess the two algorithms (FPE and CCCS) can be used for open fields, the CCCS algorithm (CCCS) for virtual wedge fields and the FPE algorithm for physical wedge fields. The dose prescription was 50 Gy in 25 fractions with photon energy of 6 MV. The RTDP were accepted such that all technique plans provided approximately a fixed level of maximal dose for optimal breast coverage while respecting the dose constraints of the organs at risk, imposed by the institution. A Siemens Artiste$^{TM}$ linac (Siemens, Munich, Germany) was used in this study. For the six RTDP techniques, the tangential fields were set up using CT-simulation, as in conventional treatment planning. The beam arrangement, including the isocenters, gantry, and collimator angles, as well as the field edges, were designed to cover the entire breast tissue with approximately a 2 cm margin in both the superior and inferior directions to provide approximately 2.5 cm beyond the skin surface anteriorly in the beam's-eye-view and perfectly align the posterior boundaries (edges) of the two tangential fields with the collimator angle set to minimize the volume of the lung in the fields (Figure 1).

The first technique was performed using the physical wedge (3DCRT-W); the second was similar but used with the Siemens virtual wedge (3DCRT-VW). For CT images in this clinical case, the suitable wedge angle was equal to 15°. For the 3DCRT-W technique, the Prowess FPE dose calculation algorithm was used (as it was the only available physical wedge algorithm calculation proposed by manufacturers), while for the 3DCRT-VW technique, the Prowess CCCS dose calculation algorithm was selected. Figure 1 shows the internal (medial) and external (lateral) tangential beam's-eye-view with the corresponding RTDP of the six techniques. For the two first techniques, only one internal tangential beam was used, and the external tangential beam was divided into two parts (segments): the first was of equal dimensions as the internal tangential beam, and the second (subfield) was closed (without the use of MLCs) in the inferior part of the breast to avoid overdose in this region, and was set not to shield the dose reference point. Therefore, initially, the dose distribution was calculated using the tangential two fields. Secondly, by viewing the dose distribution along the beam's-eye view, the subfield was optimally closed to shield the areas of the breast receiving overdoses. In this study, the weight of the closed subfield was only 5 % of the total dose. In case that an overdose region remains, an additional subfield will be necessary. According to the institution recommendations, the collimator edge of the closed beam part was paralleled to the breast motion direction to minimize respiratory motion effects. Consequently, a small MU (monitor unit) number was determined for the closed beam, only 5 % of the total dose (of the internal and external tangential); therefore, it was not possible to use the virtual wedge for this beam because the Siemens virtual wedge was not technically feasible for a small MU number. Likely due to a constraint generated by the manufacturers as well as not to exceed the dose estimation tolerance that will be disturbed when a virtual wedge beam will be delivered with a MU number approximately less than 25.

The third and fourth techniques were developed at our institution. They aim to avoid drastic changes in the doses on the edge regions of the beams, especially beam edges near the heart and lung (Figure 2). Drastic change from a high dose inside the beam (prescribed dose) to an approximately zero dose outside the beam can lead to a large difference between the estimated TPS profile and the delivered doses by the linac during the respiratory motion; this difference may be minimized by using a continuous dose degradation on the beam edges. Thus, the delivered dose during motion may be close to the planned one. Therefore, these techniques were based on the 3DCRT; they are similar to the first and second techniques, but the main changes were on the beam edges that are approximately perpendicular to the respiratory motion direction (Figure 2). This decision was made because the beam edge perpendicular to the motion direction will be the most affected, and it changes the skin projection position in the presence of motion. Conversely, the beam edges parallel to the motion direction remains approximately on the same skin projection position in the presence of motion. Consequently, the changes in the edges (Figure 2.b) of the 3DCRT-W technique inspired the new 3DCRT-W-MMT, and the changes in the edges of the 3DCRT-VW inspired the 3DCRT-VW-MMT. The beam fields generated for the continuous dose degradations are shown in Figure 1. The internal tangent was divided into three beams: the first internal tangential beam (delimited by the yellow lines in

Figure 2.b, or median beams of internal tangent of the two MMTs in Figure 1) was the same as the first two techniques (11 cm x 18.4 cm in this clinical case) but with approximately only 40 % of the entire internal tangential MU beams of the 3DCRT-W or 3DCRT-VW illustrated in Figure 2.a. Each of the other two beams (delimited by the blue and red lines in Figure 2.b) took approximately 30 % of the entire internal tangential MU beams. For the last two beams, the edges perpendicular to the motion direction of the first beam (red edges in Figure 2.b) were 1 cm closer to the beam isocenter (4.5 cm from the isocenter in this clinical case, because in Figure 2 the yellow edge was at 5.5 cm from the isocenter) and 1cm farther (6.5 cm from the isocenter) for the second beam (blue edges in Figure 2.b). The same principle was used for the external tangential beam. The only difference was that one of the two 30 % MU beams was divided into two beams (25% and 5%), where the 5 % beam hid the overdose region by closing the inferior collimator edge (as shown in Figure 1).

The fifth technique was the 3DCRT field-in-field (3DCRT-FIF) technique (Tanaka et al., 2014) using the CCCS dose calculation algorithm. The role of the static or virtual wedges in dose degradation was replaced by several MLC beams without wedges. Figure 1 shows that the first internal tangential beam was similar to that of the wedged techniques (3DCRT-W and 3DCRT-VW), but the wedge was eliminated; therefore, a second internal tangential field-in-field beam was added to compensate for the wedge role. Therefore, the overdose region was hidden by MLCs in the second internal tangential beam. The same principle was used for the external tangential beam.

The sixth technique was the direct aperture optimization IMRT (IMRT-DAO) technique (Cho, 2018; Zhang et al., 2006) using the CCCS dose calculation algorithm. This IMRT technique was developed by Zhang et al. (2006) to minimize the effects of respiratory motions for breast cancer RT. Unlike the previous conventional treatment planning for breast RT, the planning target volume (PTV) must be delineated on the CT data set to proceed with the inverse planning. The PTV was contoured to cover the clinically determined breast tissue, and superficially, it was defined as 5 mm inside of the skin surface; therefore, the lower doses in the build-up region were not considered in the optimization process. For each of the internal and external tangential beams, four segments were used. Based on the direct aperture optimization planning technique recommendations proposed by Zhang et al, 65% of the entire prescribed dose was dedicated to open internal and external flash segments (the first left IMRT-DAO segments in Figure 1) covering the entire breast and a portion of air beyond the anterior contour of the breast in the beam's-eye-view. However, the remaining 35% weight of the prescribed dose was dedicated to the last 3 segments (in each internal and external beam) to optimize the shapes and weights of each segment on the PTV, as in a classic IMRT.

*Measurements and results analyses*

A 2D dose detector, MatriXX Evolution (IBA Dosimetry, Germany), was used, which had a 1020 ionization micro chamber distributed in an active area of 24 ×24 cm2. MatriXX has been widely used for dose verification studies (Noto et al., 2014). In this experiment, it was positioned at the center of a plastic water phantom MULTICube (IBA Dosimetry, Germany) that had a length of 31.4 cm, a width of 34 cm, and a height of 22 cm (Figure 3.a).

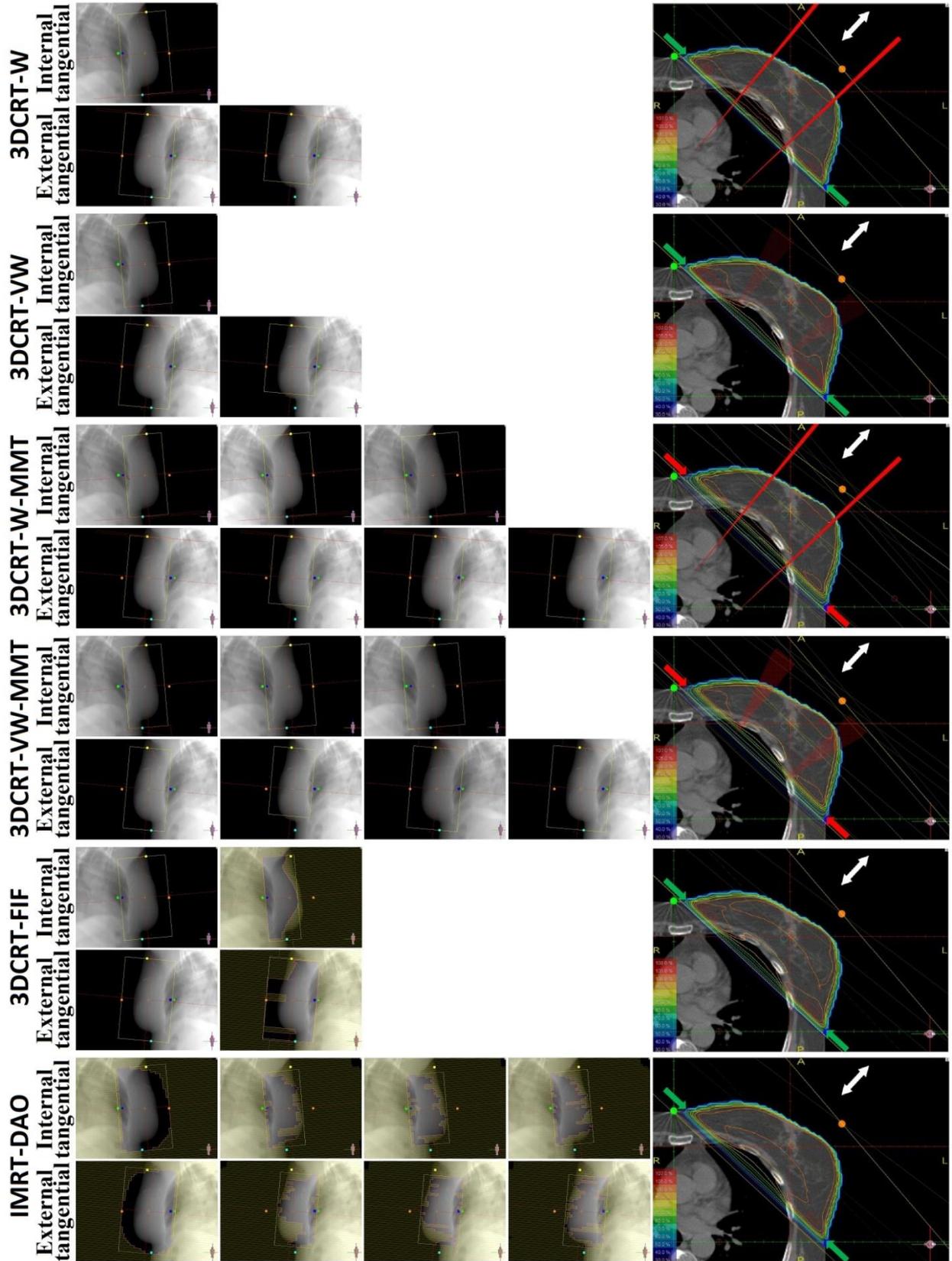

**Figure 1.** Internal and external tangential beam's-eye-view illustration for the six techniques in the left sided images, with the corresponding resultant dosimetry planning in the last right column. In the beam reference

frame, respiratory motion causes the patient anatomy to move as indicated by the white arrows. The red and green arrows indicate under-dose regions and normal dose regions in the target volume, respectively.

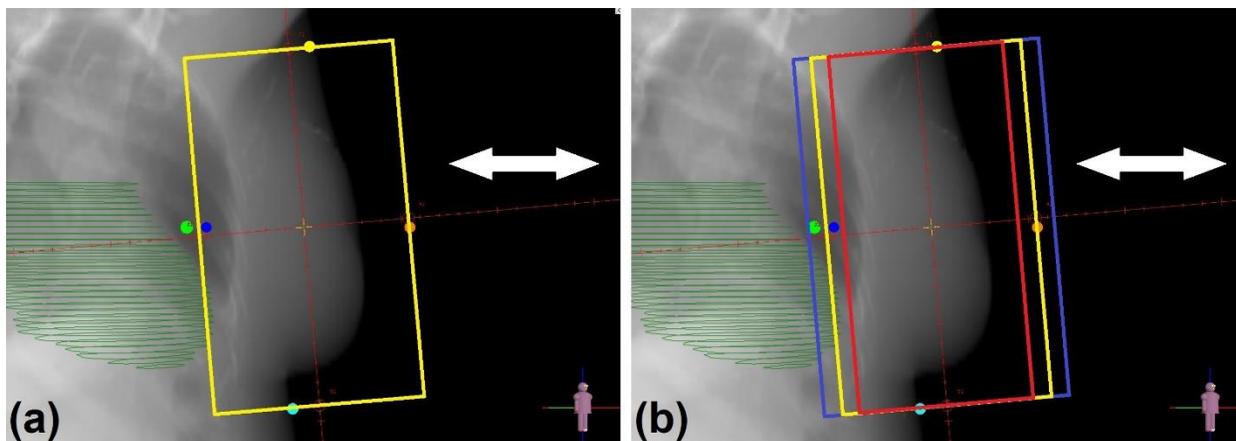

**Figure 2. An example of internal beam transformation from standard (a) to MMT (b). The internal tangent represented by yellow lines (a) was divided into three beams represented by blue, yellow and red lines (b). In the beam reference frame, respiratory motion causes the patient anatomy to move as indicated by the white arrows.**

For the six RTDP techniques, by using the TPS, dosimetry plans were exported from the breast clinical CT scan images to those of the centered MatriXX in the MULTICube. All beams and segments were centered in the center of the MULTICube, and gantry values were changed to zero degrees (Figure 3.a). MUs were conserved to be the same as those found on the breast CT scan images. After dose calculation, the 2D RT dose profiles at the 2D detectors plan position were determined from the TPS.

A sinusoidal handmade oscillator with amplitude of 2.6 cm with a period of 4 seconds was placed on the linac table (Figure 3). The oscillator was constructed by a rail included in a platform to support the translation movement (Figure 3.b). An electrical circuit and a motor attached to a metal arm were placed inside the platform (Figure 3.c). The arm was used to transform a rotational motor movement to a translation movement (Figure 3.d). The oscillator direction motion was in the Left-Right direction (shown in Figure 3.a: in the table rotation plan and perpendicular to the zero-degree angle of the table). Then, the motion direction was approximately the same as the respiratory motion if all gantry angles were changed to zero degrees (RTDP gantry angles were equal to zero degrees on the MatriXX and MULTICube). The MULTICube associated to the MatriXX was centered simultaneously in the isocenter of the linac and the center of the oscillator amplitude.

The six dosimetry plans were delivered separately on the linac, first for the internal tangent, then for the external tangent. All measurements were performed without respiratory motion simulation then with the coupled MatriXX and MULTICube oscillations. Then, the delivered doses at the MatriXX isocenter point and RT delivered 2D dose profiles were compared to those taken from the TPS. This was accomplished using the OmniPro-I'mRT (IBA Dosimetry, Germany) data analyzer by setting the gamma index criteria (Hussein et al., 2017) of 3 % / 3 mm and of 10 % / 3 mm in all the detector regions (in a ROI of 24 cm x 24 cm) for the first step, and then only 3 % / 3 mm in an inside-beam ROI of 6 cm x 6 cm centered in the isocenter.

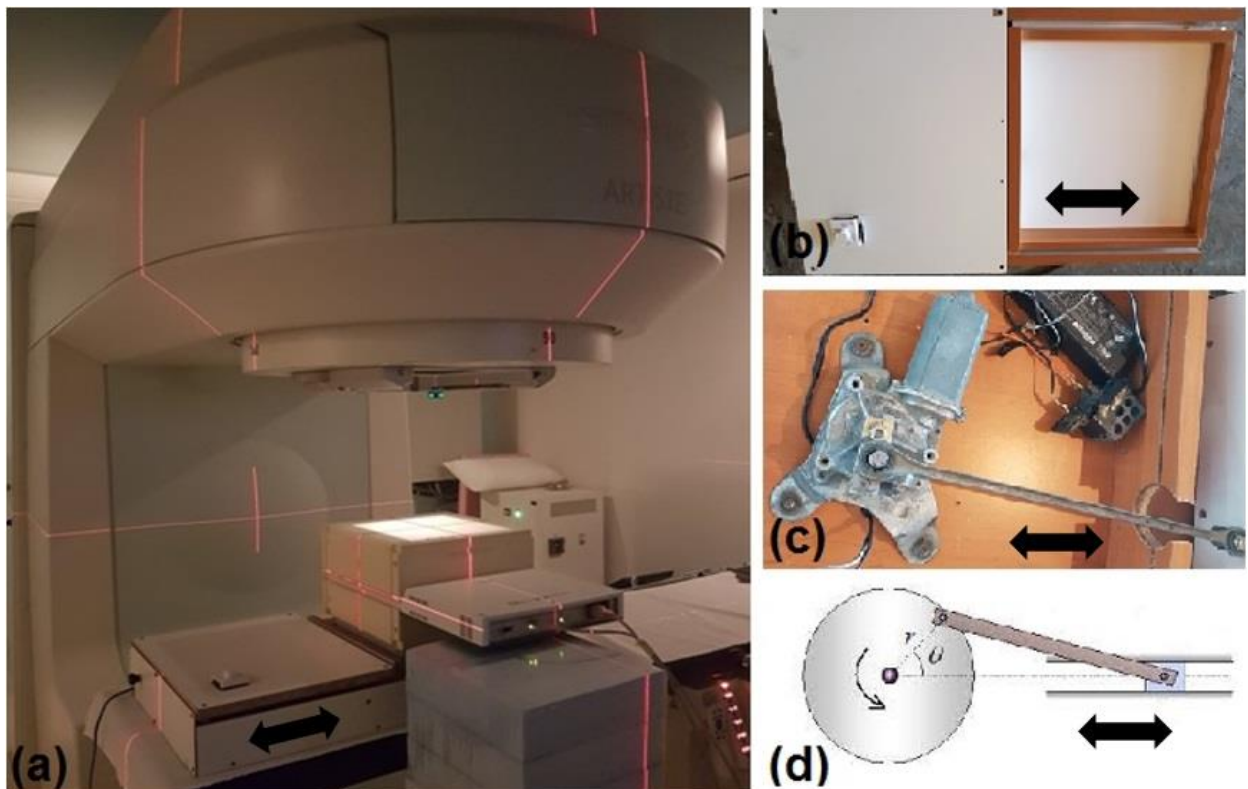

**Figure 3. MatriXX in the MULTICube, are placed on the oscillator and centered in the linac isocenter (a). Oscillator platform (b). Electrical circuit and motor with a metal arm placed inside the platform (c). Schematization of movement transformation from motor rotation to translation (d). Oscillator moves as indicated by the arrows.**

## RESULTS AND DISCUSSION

For the six used RTDP techniques, at the phantom isocenter (MatriXX and MULTICube isocenter), there were no large differences between dose values of TPS and any of the static mode in which the measurements were realized without the phantom oscillation motion or dynamic mode where the measurements were realized with the phantom oscillation motion. All differences were within ± 3 % of the TPS dose values, these results were expected only for the static mode. The surprisingly small (< 3 %) difference (rather than large) at the isocenter for the dynamic mode may be due to the simplicity of this case at the beam's-eye-view isocenter (because for this case there were no MLCs to hide the beams isocenter in all of these six techniques). However, for the six evaluated techniques, the measured dynamic isocenter doses were slightly lower than those obtained with static mode. We also note that leaves hiding the isocenter in a more complex anatomy case may lead to the opposite result.

To investigate more complex regions, Table 1 shows that in all 2D detector regions (in a ROI of 24 cm x 24 cm) with gamma index criteria of 3 % / 3 mm, the best result (high percentage) for gamma ≤ 1 was with the IMRT-DAO in static mode; this result was 96.2 % and 95.8 % for internal and external tangents respectively. Consequently, unlike the above discussed isocenter results, at least 3.8 % (100 % minus 96.2 %) of ROIs show a difference > 3 % between the planned and the delivered dose values among the six techniques. These results are schematically presented in Figures 4 and 5. These figures illustrate the measured and calculated RT dose profiles. Additionally, in the images located before the last right columns of Figures 4 and 5 (for the gamma index criteria of 3 % / 3 mm), the red color represents the region of differences (between planned and delivered doses) > 3 %. It was obvious that the red color regions were the lowest in the IMRT-DAO technique. However, in static mode, any difference between the planned (green) and delivered (red) RT dose profiles should be related only to the precision of the TPS dose calculation algorithm or linac commissioning. Therefore, the lowest gamma index % was for the two physical wedge techniques (56.2 %, 58.4 %, 57.8 % and 58.9 %, respectively for internal and external tangents of 3DCRT-W and internal and external tangents of 3DCRT-W-MMT, most probably because the used Prowess FPE dose calculation algorithm was not suitable for scattered photons dose calculation outside the beam field. As illustrated in Figures 4 and 5, for 3DCRT-W and 3DCRT-W-MMT, the green (planned) RT dose profiles were at zero values approximately at more than 1.5 cm outside the beam boundary (outside X = ± 7 cm and Y = ± 11 cm; note that the beam field size was X = ± 5.5 cm and Y = ± 9.2 cm for this clinical case). A difference in low doses values (largely outside the edges of the fields) between the planned (green) and the delivered (red) profiles was caused by these zero values. Consequently, in these two techniques and with gamma index criteria of 3 % / 3 mm, Figures 4 and 5 showed that the red color covered the area approximately at more than 1.5 cm outside the beam field. However, Figures 4 and 5 showed that in static mode, for all the six used techniques, the doses were well estimated inside the fields and approximately at less than 1.5 cm outside the edges of the fields (the blue color covered the area). Consequently, in static mode, concerning the precision of the TPS dose calculation algorithm and linac commissioning, all the six used techniques were eligible for clinical use in radiotherapy where high treatment dose values are used (in Gy) and the low doses located far from the beam fields can be ignored. On the other hand, the CCCS dose calculation algorithm, which is suitable for scattered photons dose calculation, was used for the virtual wedge and the FIF techniques. For this reason, in Figures 4 and 5, the green (planned) RT dose profiles where simulated (non-zero values) at far regions outside the beam field, and the gamma index % were increased for the CCCS dose calculation algorithm techniques (88.4 %, 87.7 %, 93.2 %, 92.4 %, 90.9 % and 90 %, respectively for internal and external tangents of 3DCRT-VW, internal and external tangents of 3DCRT-VW-MMT and internal and external tangents of 3DCRT-FIF). Red regions of low doses remain at approximately 1.5 cm outside the beam edges. These low doses estimation errors may be due to a lack of consideration of low scattered dose outside the beam field during linac commissioning. These low dose errors were reduced for IMRT technique, likely due to an improvement in the process during IMRT linac commissioning. On the other hand, for the gamma index criteria of 10 % / 3 mm, approximately 100 % of the gamma index was ≤ 1 for the six techniques in the static mode. Consequently, there was no difference > 10 % from the TPS planned doses. Therefore, for these six techniques, the error of the dose estimation related to the precision of the TPS dose calculation algorithm and linac commissioning was < 10 % in all regions (outside and inside the beams fields projections), and this result is schematically presented by the disappearance of the red color in Figures 4 and 5 for the gamma index criteria of 10 % / 3mm.

**Table 1. Percentages of gamma values less than or equal to 1, with the corresponding maximum gamma signal values for the six techniques.**

| | | Gamma information | | | | Radiotherapy planning techniques | | | | | |
|---|---|---|---|---|---|---|---|---|---|---|---|
| | | Gamma ROI (cm x cm) | Gamma % | Gamma mm | | 3DCRT-W | 3DCRT-VW | 3DCRT-W-MMT | 3DCRT-VW-MMT | 3DCRT-FIF | IMRT-DAO |
| Internal tangential | Without motion | 24 x 24 | 3 | 3 | % gamma ≤ 1 : | 56.2 | 88.4 | 57.8 | 93.2 | 90.9 | 96.2 |
| | | 24 x 24 | 3 | 3 | Max. gamma : | > 2 | 1.5 | > 2 | 1.5 | 1.6 | 1.4 |
| | | 24 x 24 | 10 | 3 | % gamma ≤ 1 : | 99.8 | 100 | 100 | 100 | 100 | 100 |
| | | 24 x 24 | 10 | 3 | Max. gamma : | 1.1 | 0.9 | 0.9 | 0.9 | 0.9 | 0.8 |
| | | 6 x 6 | 3 | 3 | % gamma ≤ 1 : | 100 | 100 | 100 | 100 | 100 | 100 |
| | | 6 x 6 | 3 | 3 | Max. gamma : | 0.3 | 0.5 | 0.3 | 0.4 | 0.6 | 0.5 |
| | With motion | 24 x 24 | 3 | 3 | % gamma ≤ 1 : | 39.5 | 74.3 | 47.5 | 77.5 | 75.9 | 81.3 |
| | | 24 x 24 | 3 | 3 | Max. gamma : | > 2 | > 2 | > 2 | > 2 | > 2 | > 2 |
| | | 24 x 24 | 10 | 3 | % gamma ≤ 1 : | 90.7 | 89.9 | 99.1 | 100 | 92.5 | 94.2 |
| | | 24 x 24 | 10 | 3 | Max. gamma : | 1.7 | 1.9 | 1.1 | 1 | 1.9 | 1.7 |
| | | 6 x 6 | 3 | 3 | % gamma ≤ 1 : | 100 | 100 | 100 | 100 | 100 | 99.6 |
| | | 6 x 6 | 3 | 3 | Max. gamma : | 0.4 | 0.9 | 0.9 | 0.8 | 1 | 1.1 |
| External tangential | Without motion | 24 x 24 | 3 | 3 | % gamma ≤ 1 : | 58.4 | 87.7 | 58.9 | 92.4 | 90.0 | 95.8 |
| | | 24 x 24 | 3 | 3 | Max. gamma : | > 2 | 1.5 | > 2 | 1.5 | 1.5 | 1.4 |
| | | 24 x 24 | 10 | 3 | % gamma ≤ 1 : | 100 | 100 | 100 | 100 | 100 | 100 |
| | | 24 x 24 | 10 | 3 | Max. gamma : | 1 | 0.8 | 0.9 | 0.9 | 0.8 | 0.8 |
| | | 6 x 6 | 3 | 3 | % gamma ≤ 1 : | 100 | 100 | 100 | 100 | 100 | 100 |
| | | 6 x 6 | 3 | 3 | Max. gamma : | 0.5 | 0.6 | 0.5 | 0.6 | 0.6 | 0.7 |
| | With motion | 24 x 24 | 3 | 3 | % gamma ≤ 1 : | 41.8 | 74.5 | 50.3 | 78.1 | 74.9 | 79.4 |
| | | 24 x 24 | 3 | 3 | Max. gamma : | > 2 | > 2 | > 2 | > 2 | > 2 | > 2 |
| | | 24 x 24 | 10 | 3 | % gamma ≤ 1 : | 91.4 | 90.8 | 100 | 99.4 | 91.8 | 94.1 |
| | | 24 x 24 | 10 | 3 | Max. gamma : | 1.6 | 1.8 | 1.1 | 1.1 | 1.7 | 1.5 |
| | | 6 x 6 | 3 | 3 | % gamma ≤ 1 : | 100 | 100 | 100 | 100 | 98.2 | 87.9 |
| | | 6 x 6 | 3 | 3 | Max. gamma : | 0.5 | 0.6 | 0.4 | 1 | 1.3 | 1.4 |

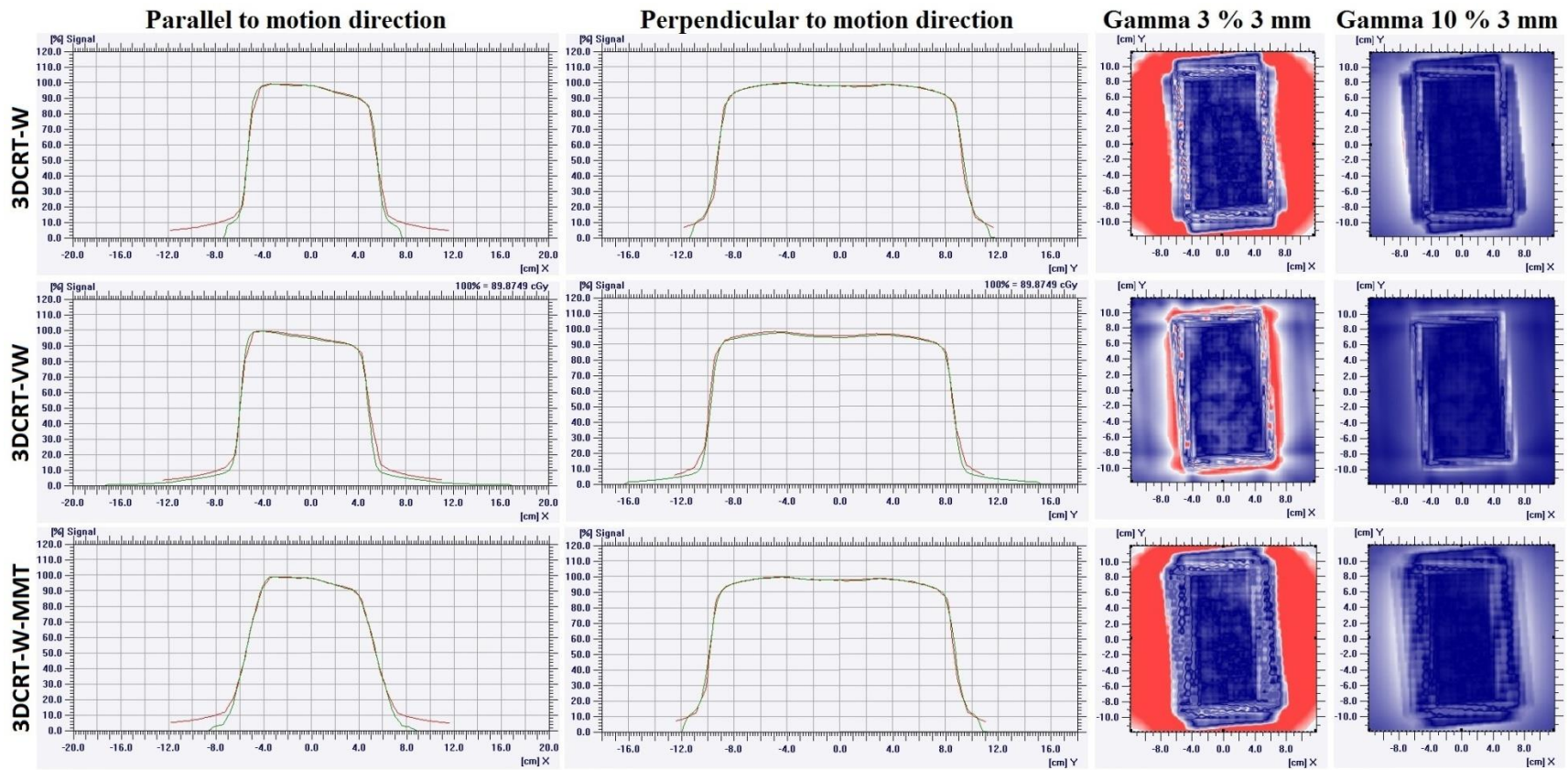

**Figure 4.** From left to right: the planned (green) and delivered (red) internal tangential RT dose profiles at the isocenter in static mode for the parallel motion direction (X, in the left column) followed by the perpendicular motion direction (Y, in the second column), correspondent to the 2D region (X, Y) showing the gamma values for index criterion of 3 % / 3 mm followed by the index criterion of 10 % / 3 mm. Each image line corresponds to a different technique. The red spots in the last two columns highlight the regions of the gamma values (levels in the image) greater than 1 (outside index criteria tolerance). Arrows illustrate the direction of the motion simulation.

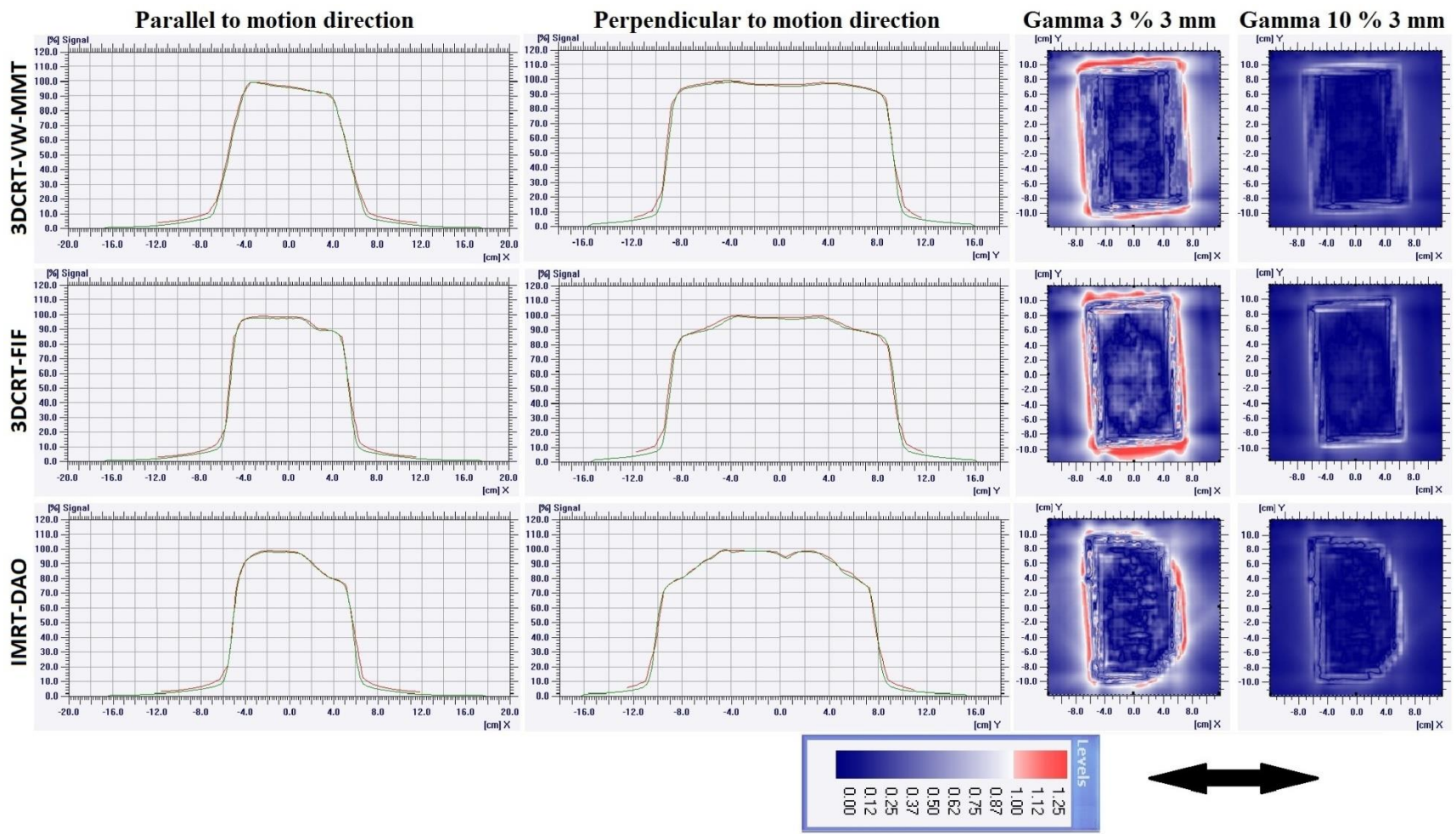

**Figure 4. (continued)**

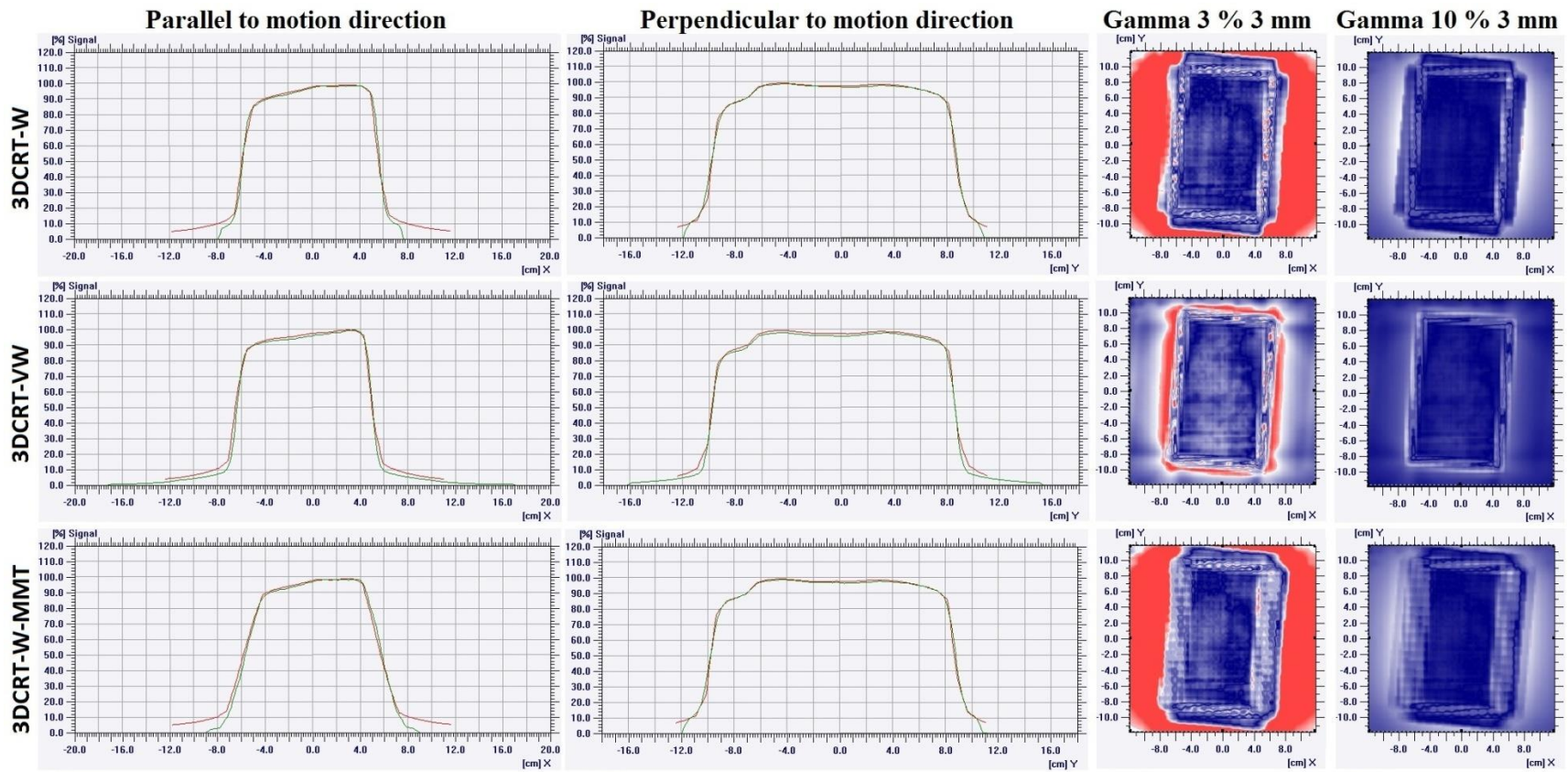

Figure 5. From left to right: the planned (green) and delivered (red) external tangential RT dose profiles at the isocenter in the static mode for the parallel motion direction (X, in the left column) followed by the perpendicular motion direction (Y, in the second column), correspondent to the 2D region (X, Y) showing the gamma values for the index criterion of 3 % / 3 mm followed by the index criterion of 10 % / 3 mm. Each image line corresponds to a different technique. The red spots in the last two columns highlight the regions of gamma values (levels in the image) > 1 (outside index criteria tolerance). Arrows illustrate the direction of the motion simulation.

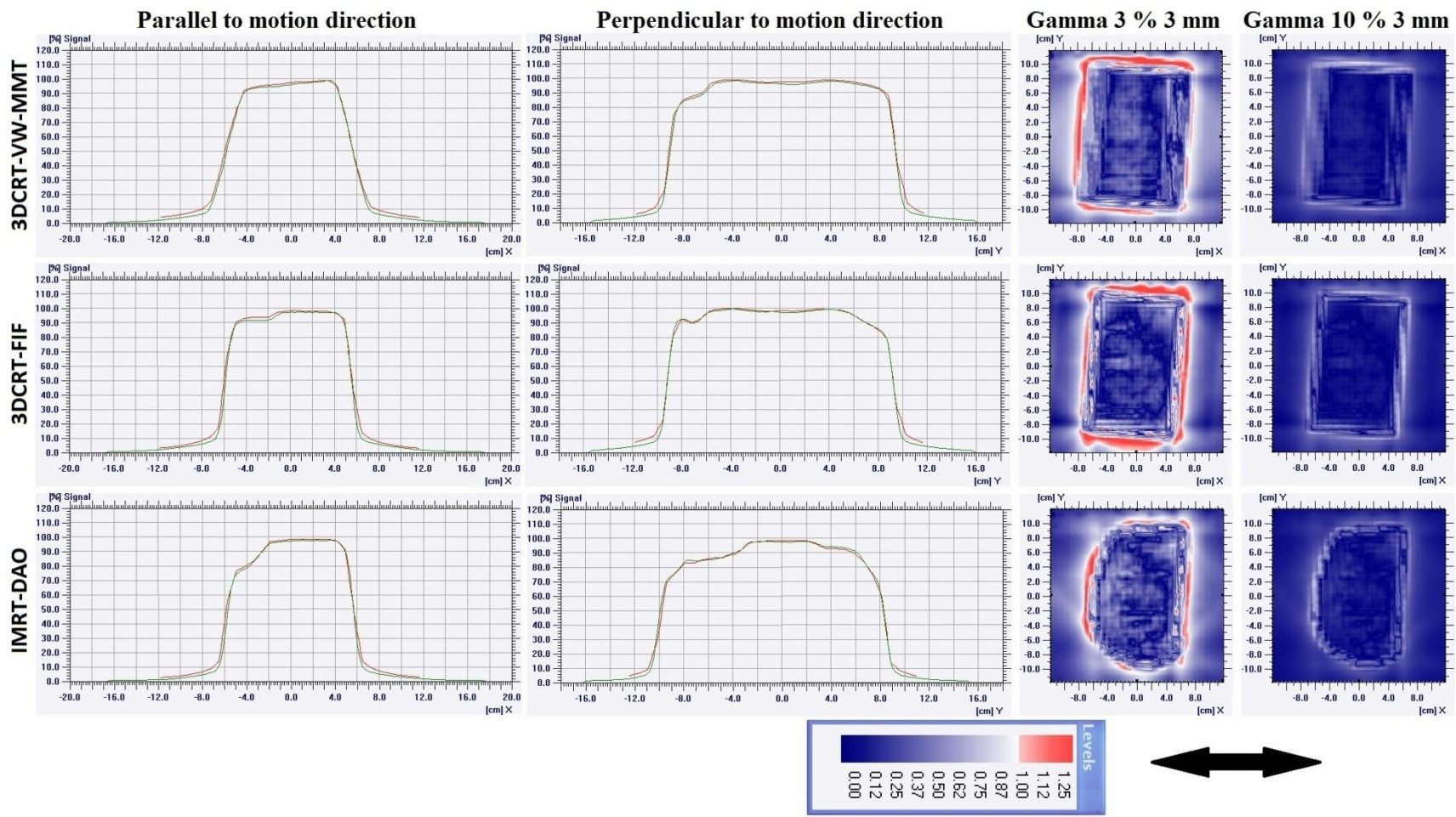

**Figure 5. (continued)**

In the presence of motion, Table 1 shows that in the large ROI (24 cm x 24 cm) for gamma index criteria of 3 % / 3 mm (red color in Figures 6 and 7 for 3 % / 3 mm), the gamma index % was degraded widely by leading gamma index maximal values superior to 2 for all techniques. As described by Hussein et al. (2017), the gamma index is calculated based on finding the minimum Euclidean distance for each reference point. Euclidean distance depends on the distance between reference to evaluated point, dose difference between reference to evaluated point and gamma index criteria. Therefore, in a particular condition, when the distance between reference to evaluated point is equal to 0 mm: the difference between reference to evaluated point should be greater or equal to the value of gamma index in this reference point, multiplied by the % of gamma index dose criterion. Therefore, for the previous gamma index maximal values superior to 2 (the maximum value of gamma index that can be displayed by the used data analyzer was equal to 2) in the six techniques (Table 1), motion leads to the appearance of regions with differences superior to 6 % (2 x 3 %) from the calculated estimated TPS doses. Therefore, in this motion situation, by using gamma index criteria of 3 % / 3 mm the six techniques cannot be compared because the only information was that the differences for all techniques was superior to 6 % but there was not an exact value for these differences. Consequently, gamma index criteria of 10 % / 3 mm was used in order to be able to compare techniques and to enlarge the displayed differences value from 6 % to 20 % (2 x 10 %). In Figures 6 and 7, for gamma index criteria of 10 % / 3 mm, the red color is greatly decreased only for the proposed MMTs (3DCRT-W-MMT and 3DCRT-VW-MMT). Consequently, on these clinical case dataset images, for these two RTDP realized by MMT (3DCRT-W-MMT and 3DCRT-VW-MMT), the motion affected the planned doses less than 10 % in more than 99.1% of regions (Table 1), and with a maximum difference of only 11 % (10 % x Max. gamma of Table 1) in the remaining 0.9 % region (100 % minus 99.1%). Nevertheless, in the other four techniques, motion leads to a dose difference between 15 to 19 % (10 % x Max. gamma of Table 1) from the planned dose (for the 19 %, gamma pass rates in all these regions were 89.9 % for 10 % / 3 mm). As shown in Figures 6 and 7, for the gamma index criteria of 10 % / 3 mm, the red color was located in a critical area on the lateral edges of the beams; thus, this error of estimation can eventually lead to an unexpected overdose to the patient. In addition, it was clear in Table 1 that for gamma index criteria of 3 % / 3 mm, during the phantom motion the pass rates were similarly reduced for MMTs and for techniques without motion management (for example, for internal tangential, reductions were 14.1 % for 3DCRT-VW and also about 15.7 % for 3DCRT-VW-MMT). This is not the case for gamma index criteria of 10 % / 3mm when large effects of MMTs where appeared (for example, for internal tangential, reductions were 10.1 % for 3DCRT-VW and 0 % for 3DCRT-VW-MMT). In clinical situation, it was recommended to use gamma criteria values near to 3 % / 3 mm (Hussein et al., 2017). These low dose values of gamma criteria (3 % / 3 mm) were used generally in static mode during the pretreatment QA without taking into account the motion in real time dose delivery. However, in the current study we were interested largely for a high value of gamma criteria (10 % / 3 mm), because we were aiming to detect a high over or under dose that may occurs during patient treatments without our knowledge. For the four techniques used in this study without motion management, the radiation oncologists commonly validate the dosimetry plan without considering the real distribution of the radiation dose in the presence of respiratory motion. By using a high value of gamma criteria (10 % / 3 mm), our study shows that for breast patient with a large respiratory amplitude near to 2.6 cm, the real delivered dose distribution may be largely deviated from the planned one, and can eventually lead to an unexpected overdose to the patient as mentioned previously.

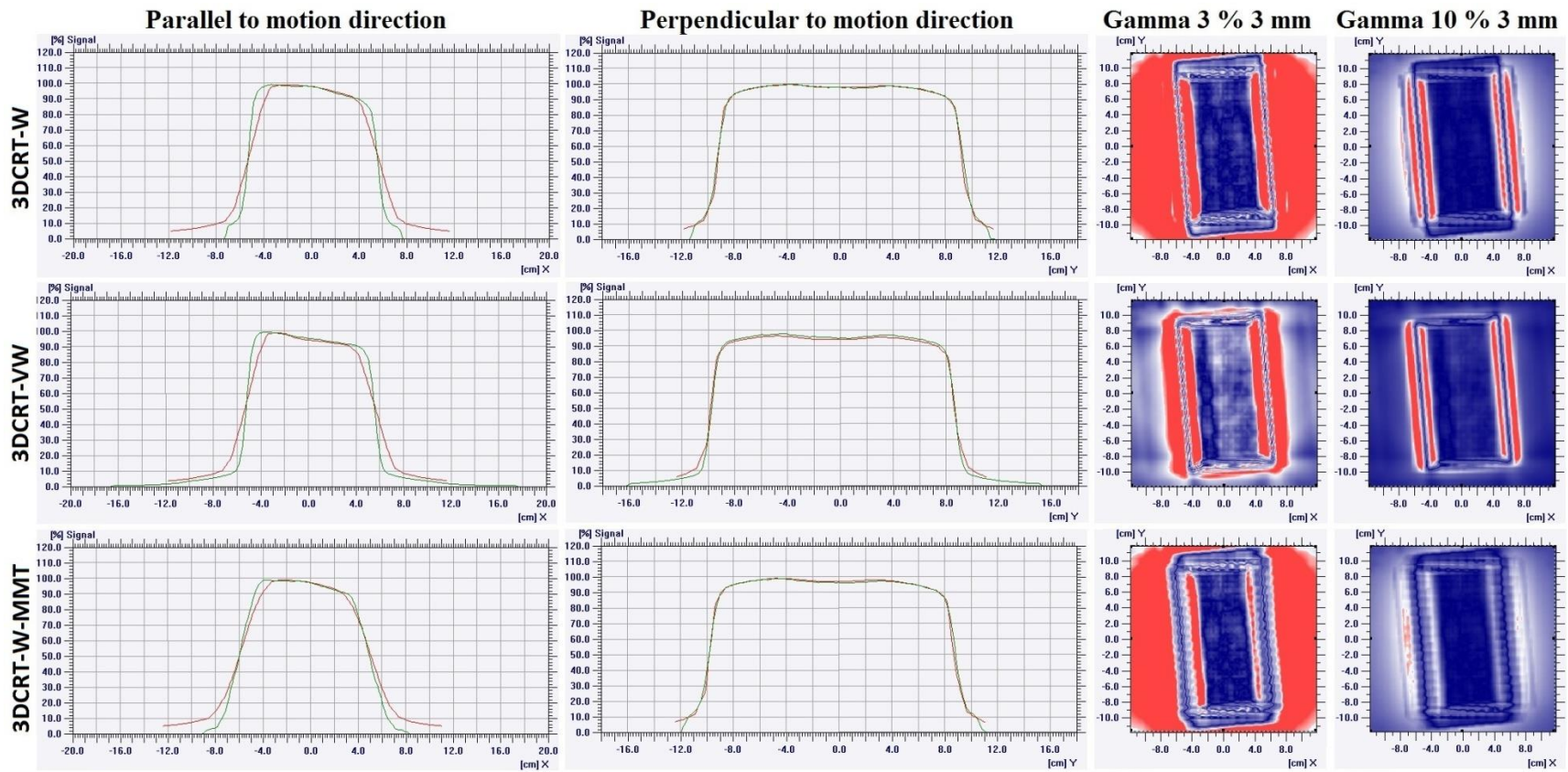

Figure 6. From left to right: the planned (green) and delivered (red) internal tangential RT dose profiles at the isocenter in the dynamic mode for the parallel motion direction (X, in the left column) followed by the perpendicular motion direction (Y, in the second column), correspondent to the 2D region (X, Y) showing the gamma values for the index criterion of 3 % / 3 mm followed by the index criterion of 10 % / 3 mm. Each image line corresponds to a different technique. The red spots in the last two columns highlight the regions of gamma values (levels in the image) > 1 (outside index criteria tolerance). Arrows illustrate the direction of the motion simulation.

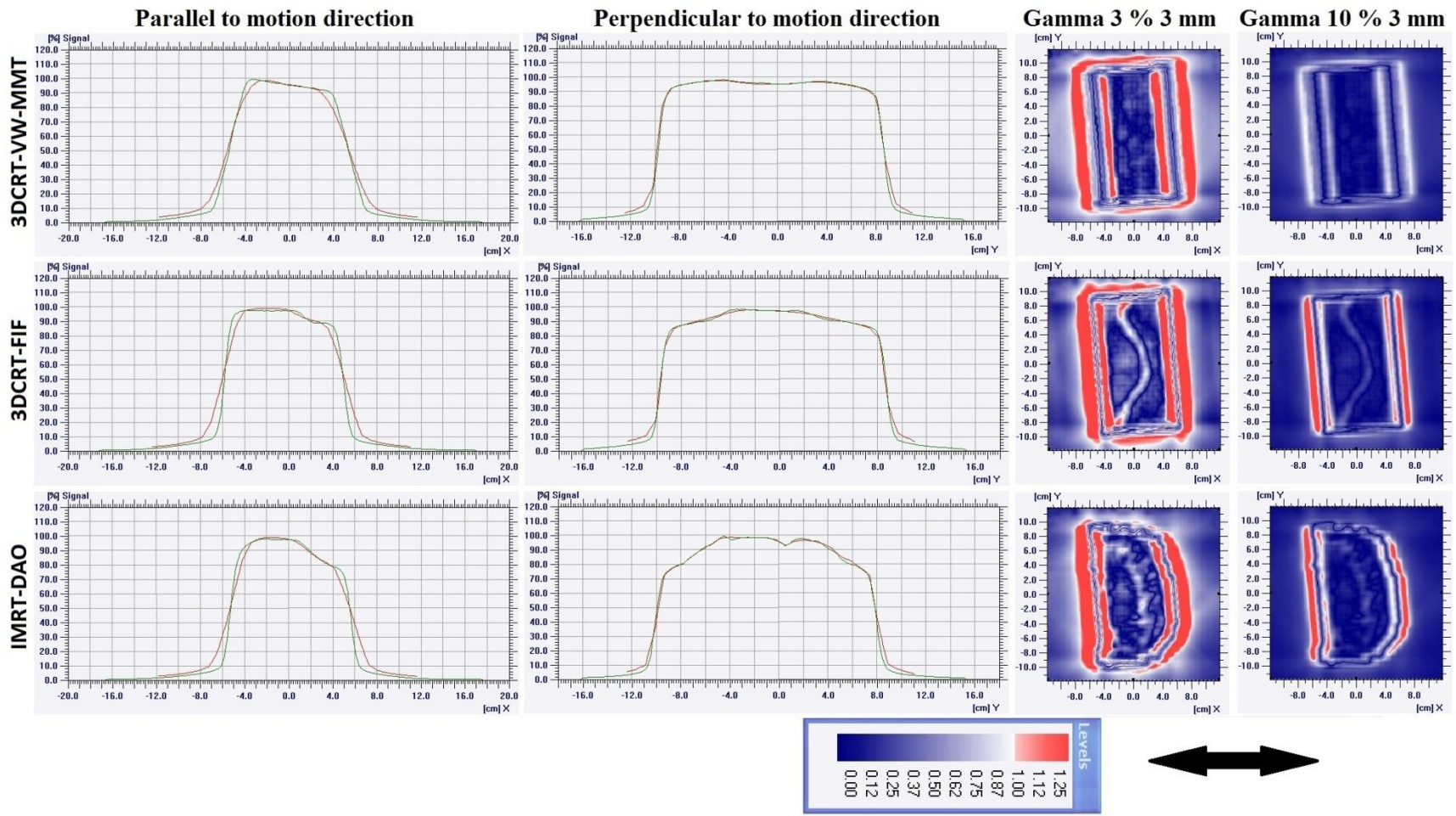

**Figure 6. (continued)**

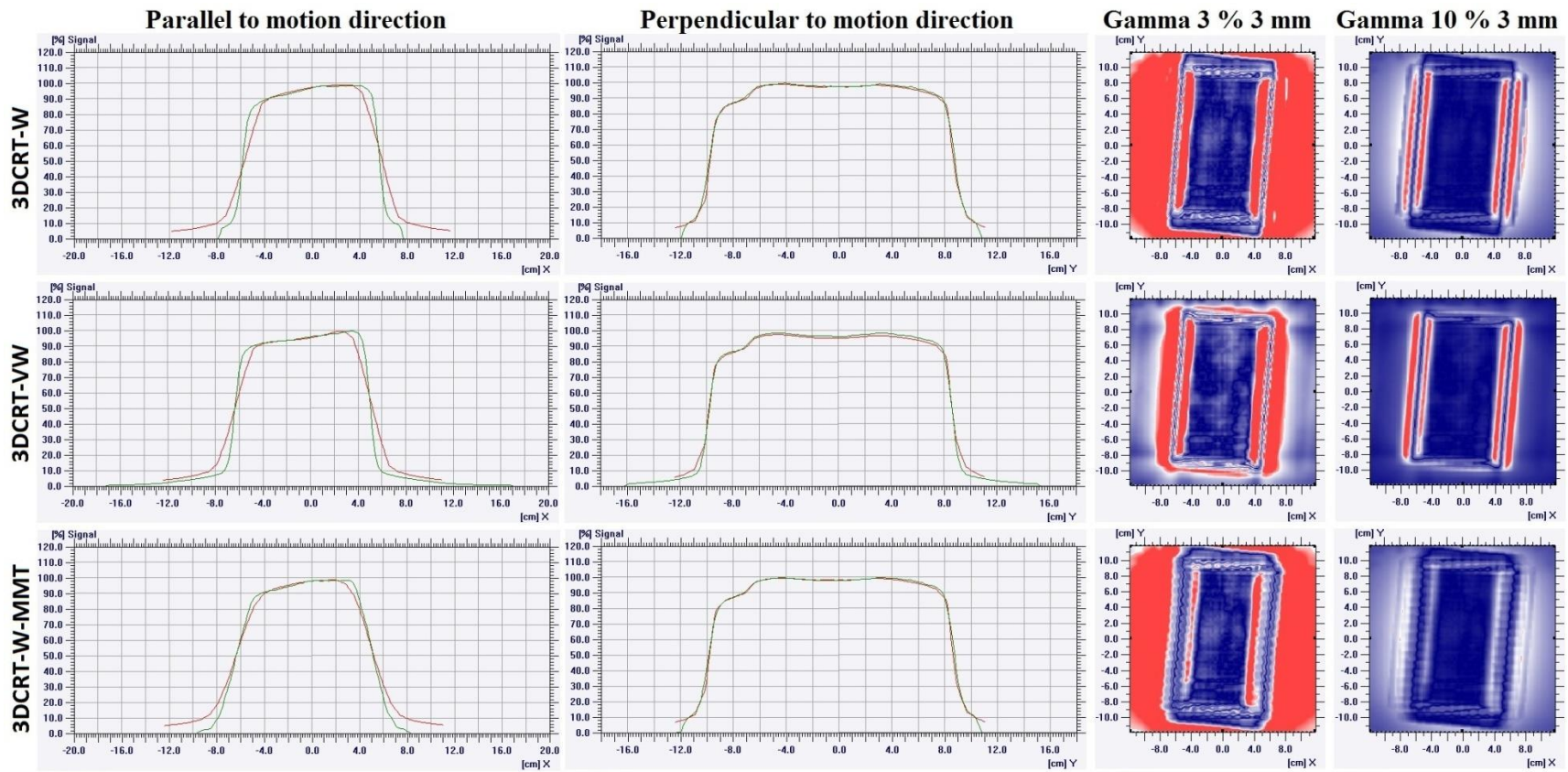

Figure 7. From left to right: the planned (green) and delivered (red) external tangential RT dose profiles at the isocenter in the dynamic mode for the parallel motion direction (X, in the left column) followed by the perpendicular motion direction (Y, in the second column), correspondent to the 2D region (X, Y) showing the gamma values for the index criterion of 3 % / 3 mm followed by the index criterion of 10 % / 3 mm. Each image line corresponds to a different technique. The red spots in the last two columns highlight the regions of gamma values (levels in the image) > 1 (outside index criteria tolerance). Arrows illustrate the direction of the motion simulation.

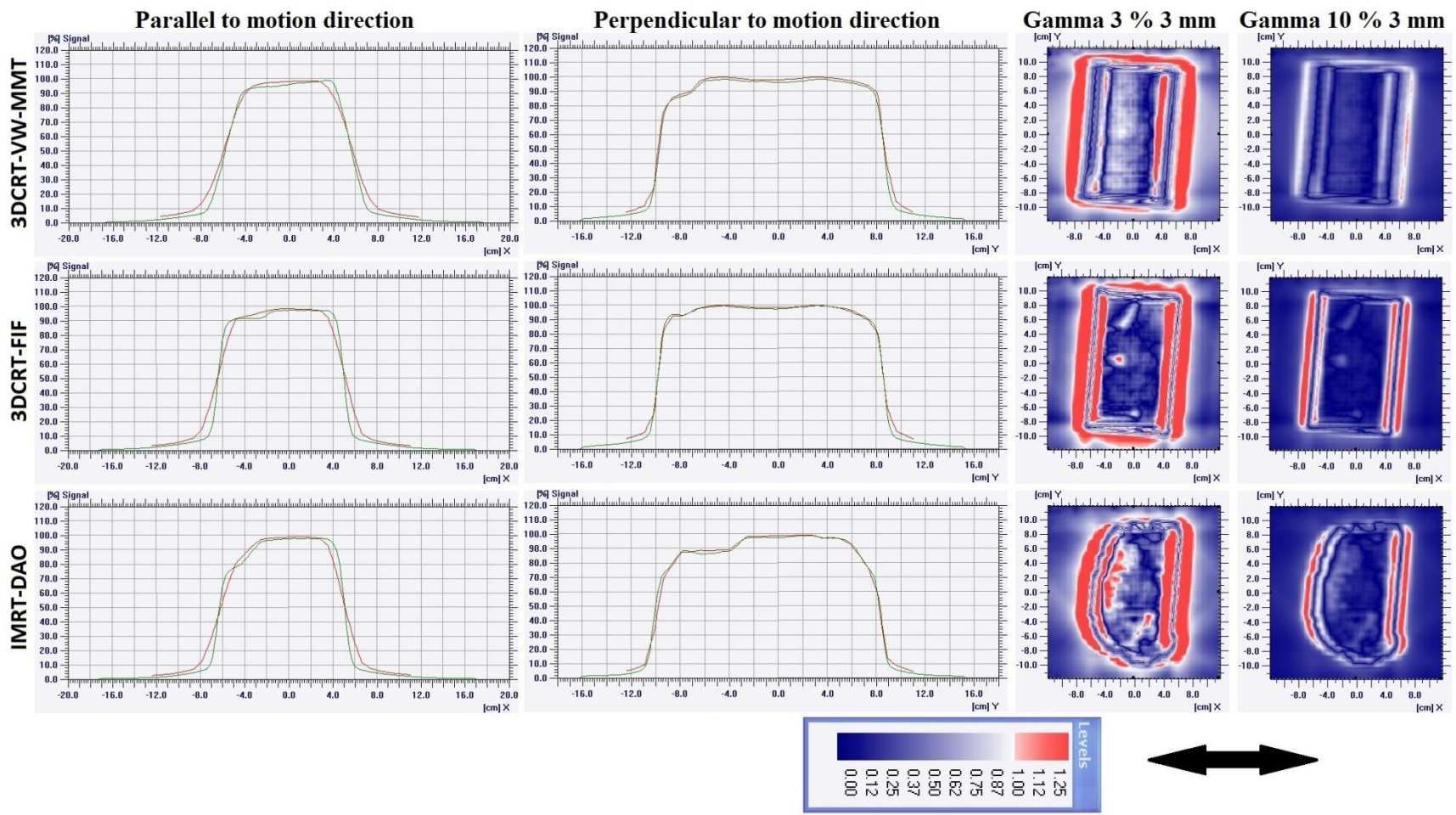

Figure 7. (continued)

On the other hand, to isolate the high effect of motion in beam edge regions, Table 1 showed the measurement results for the gamma index criteria of 3 % / 3 mm inside an ROI of 6 cm x 6 cm centered in the isocenter. Also, it showed that in the static mode, there was no difference between the planned and the measured doses for the six techniques. However, in the dynamic mode, the dose estimation was degraded for the techniques in which MLCs were used (3DCRT-FIF and IMRT-DAO): from a difference less than 3 % (gamma pass rates were 100 % for 3 % / 3 mm) to a difference up to 3 % (for the external tangential of IMRT-DAO technique, gamma pass rates were 87.9 % for 3 % / 3 mm). For example, for the gamma index criteria of 3 % / 3 mm presented in Figure 7 in which an overdose red spot was located near the center for the 3DCRT-FIF technique. This spot was associated with MLCs, which are used in Figure 1 to hide the overdose on the TPS by using another external tangential beam. In Table 1, this red spot is associated with a value of 98.2 % for a "gamma index ≤ 1". Therefore, 1.8 % (100 % minus 98.2 %) of the ROIs were outside the index criteria tolerance (difference > 3 % in 1.8 % of the ROI). Consequently, if motion changes MLCs projections positions on the patient anatomy, then an unexpected overdose will occur.

It was obvious from the beginning of the study that these MMTs suffer from a main disadvantage: the dose blurring at the beam edge regions (continuous doses degradation) resulting in regions with unwanted under-dose in the target volume and with over-dose in normal tissue. On the dosimetry plans images of Figure 1, the red arrows showed the unwanted under-doses regions in the target for the MMTs. Also, the unwanted under and over doses were seen in the continuous dose degradation regions of the planned (green color) and delivered (red color) RT profiles involving the static and dynamic modes for MMTs as shown in Figures 4, 5, 6 and 7. On the other hand, for the techniques when motion management was not used, the unwanted under and over doses were present just in the delivered RT profile in the dynamic mode only as shown in Figures 6 and 7. However, for these techniques used without motion management, the unwanted doses were also present in real time treatment but although not seen on the TPS. In addition, if these MMTs will be applied in the future, we hope that radiation oncologists will accept the dosimetry plans involving these unwanted under-doses reaching the target regions (Figure 1), based on continuous study which prove that these doses are acceptable on the peripheral region of the breast clinical target volume (CTV) without causing tumor recurrence in the future. Otherwise, recent study shows that some regions in breasts can tolerate high maximum doses, they can achieve in some cases approximately 112 % of the prescribed 50 Gy, without causing any skin dermatitis (Moussallem et al., 2019). Consequently, the unwanted under-doses maybe can be partially eliminated by increasing the dose in the target without causing any skin damage.

Furthermore, it is interesting to develop and try a non-tangential MMT based on dose volumes degradation instead of beams eye's view degradations (like non-tangential IMRT-MMT), by developing a similar MMT for a continuous dose level degradation from the CTV to the PTV throughout the creation of several PTVs with different dose prescription values. Thus, the severity of dose degradation should be a compromise between the under-doses reaching the PTV and the efficiency of the MMT.

Finally, the principal aim of this study was to test if these MMTs are usefully and acceptably applicable. The previously found results have opened the door to go further in working on detailed studies that firstly aim to improve the efficiency of the MMTs by studying the effects of respiration motion amplitudes and select the corresponding parameters for each: beam size of degradation (1 cm in the current study) and beams weighting of dose percentages (30 %, 40 % and 30 % in the current study), and secondly to enhance its validation. Actually, using a 2D gamma index evaluation (in this study) doesn't give exact geometric location of the dose differences. The red regions on the gamma map are not necessarily on the heart and lung. They give an idea of the correlations between calculated and measured dose in one particular verification plan. A volumetric evaluation (3D gamma index map) with a measured DVH (dose-volume histogram) are needed in order to accurately evaluate the dose difference on the heart and lung. On the other hand, the study might be extended to more patients (more CT images sets. Not only on 1 CT images set) for better statistical significance. At the final validation step, measurements of the delivered doses should be done by using in-vivo dosimetry on anthropomorphic phantoms or real patients rather than the classics phantoms like that was used in this study.

# CONCLUSION

In breast radiation therapy, the radiation oncologist widely validates a "best" dosimetry plan without knowing the real distribution of the radiation dose in the presence of respiratory motion. In fact, the real dose distribution may be largely affected by the motion and was shown to be largely different from the planned one. In tangential techniques, depending on the breast anatomy, the motion can affect the dose estimation by more than 3 %, due to the use of MLCs, and 19 %, due to the discontinuity of dose on the beam edges. However, a new motion management technique for breast radiation therapy dosimetry planning was proposed based on the continuous dose degradation and without the use of MLCs. Initial promising results were found by using the motion management technique, finding that the radiation dose estimation was affected by approximately 11 % on the beam edges and < 3 % inside the beams projections. So, the main advantage of this new dosimetry planning is reducing the motion effect to estimate the delivered dose accurately.

# ACKNOWLEDGMENTS

Authors would like to thank Dr. Bassam Hajj, Dr. Maurice Moarbes and Ms. Maya Dandachi for editing the paper, the "Centre de Traitement Médical du nord" team for their support, as well as Mr. Georges Greige for his contribution in the development of the respiratory motion oscillator.